\documentclass[a4paper,fleqn]{cas-sc}

\usepackage[authoryear,longnamesfirst]{natbib}
\usepackage[utf8]{inputenc}
\usepackage{booktabs}
\usepackage{array}
\usepackage{tabularx} 
\def\tsc#1{\csdef{#1}{\textsc{\lowercase{#1}}\xspace}}
\tsc{WGM}
\tsc{QE}
\tsc{EP}
\tsc{PMS}
\tsc{BEC}
\tsc{DE}

\begin{document}
\let\WriteBookmarks\relax
\def\floatpagepagefraction{1}
\def\textpagefraction{.001}

\shorttitle{Qualitative and quantitative analysis of student's perceptions in the use of generative AI in educational environments}

\shortauthors{S. Altares-López et~al.}

\title [mode = title]{Qualitative and quantitative analysis of student's perceptions in the use of generative AI in educational environments}
\author[1]{Sergio Altares-López}[type=author,
                        auid=000,bioid=1,
                        orcid=0000-0002-0847-6113]

\ead{sergio.altares@csic.es}



\affiliation[label1]{organization={Consejo Superior de Investigaciones Científicas, Centre for Automation and Robotics-CAR (CSIC-UPM)},
            addressline={Ctra. Campo Real km. 0,200},
            postcode={28500},
            state={Arganda},
            country={Spain}}

\affiliation[label2]{organization={Department of Electrical, Electronical and Automatic Control Engineering and Applied Physics, Escuela Técnica Superior de Ingeniería y Diseño Industrial, Universidad Politécnica de Madrid},postcode={28012}, state={Madrid},country={Spain}}

\affiliation[label3]{organization={Centro de Investigación e Innovación Eléctrica, Mecánica y de la Industria (CINEMI), Universidad Tecnológica de Panamá},
            country={Panamá}}

\author[1]{José M. Bengochea-Guevara}[style=chinese, orcid = 0000-0003-4081-7325]
\ead{jose.bengochea@csic.es}

\author[1]{Carlos Ranz}[style=chinese]
\ead{carlos.ranz@car.upm-csic.es}

\author[2,3]{Héctor Montes}[style=chinese, orcid= 0000-0001-8638-9966]
\ead{hector.montes1@utp.ac.pa}

\author[1]{Angela Ribeiro}[style=chinese, orcid= 0000-0001-5807-8132]
\ead{angela.ribeiro@csic.es}

\cormark[1]


\credit{Data curation, Writing - Original draft preparation}

\begin{abstract}
The effective integration of generative artificial intelligence in education is a fundamental aspect to prepare future generations. The objective of this study is to analyze from a quantitative and qualitative point of view the perception of controlled student-IA interaction within the classroom. This analysis includes assessing the ethical implications and everyday use of AI tools, as well as understanding whether AI tools encourage students to pursue STEM careers. Several points for improvement in education are found, such as the challenge of getting teachers to engage with new technologies and adapt their methods in all subjects, not just those related to technologies.
\end{abstract}



\begin{keywords}
generative AI \sep  student's perception \sep  learning science \sep human-centered artificial intelligence \sep AI education
\end{keywords}

\maketitle

\section{Introduction}

Education has undergone significant transformation in recent decades, largely driven by technological advancements. In this context, artificial intelligence (AI) has emerged as a promising tool with the potential to revolutionize teaching and learning methods \citep{zaman2023transforming,huang2021review}. As AI continues to evolve and become increasingly integrated into society, there is a growing need to educate the next generation about its capabilities, ethical implications, and potential for both innovation and misuse. One of the most innovative areas within AI is the so-called \textit{generative AI}, which refers to systems capable of creating new and original content. These systems can generate texts with large language models (LLMs) \citep{achiam2023gpt, touvron2023llama, team2023gemini}, images using diffusion models, generative adversarial networks (GANs) or variational autoencoders (VAEs) \citep{salimans2016improved,borji2022generated}, music by using wavenet or other transformers architectures \citep{dong2018musegan,briot2017deep, wang2024review} and other types of data that can be used in a variety of applications.

The application of AI in education offers unique opportunities to personalize learning, adapting to the individual needs of each student \citep{pesovski2024generative}, so they are systems that prioritize human needs, values and experiences, in this case of students and teachers. Different studies have been conducted to bring AI closer to educational centres at different levels and workflows \citep{holmes2022state, baidoo2023education,lim2023generative,chiu2023impact,alasadi2023generative,sharples2023towards,ruiz2023empowering,morch2023human,yang2021human,yang2023guest,cao2024survey,ruiz2023empowering}. In addition, it allows educators to create interactive and dynamic educational materials that can increase student participation and engagement. However, effectively integrating generative AI into the classroom presents significant challenges, both technical and ethical \citep{alasadi2023generative}.

However, introducing AI concepts to students with no prior knowledge can be a complex task, which would require a lot of time and resources. Additionally, the teaching staff will need to be prepared to adapt the content of their classes with AI techniques, which is an even greater challenge \citep{liua2021artificial}. In particular, this research aims to address this challenge by presenting a novel methodology tailored to introduce students to the world of AI-generated videos in an intensive session. In this way, students not only gain hands-on experience in AI creation but also develop a fundamental understanding of the risks and liabilities associated with AI technologies. In addition, this approach is designed to spark interest and curiosity, encouraging students to consider pursuing careers in AI-related fields \citep{zastudil2023generative,tossell2024student,ngo2023perception,vallis2023student,leiker2023generative,țalua2024exploring,su2023unlocking,vartiainen2023using}.

The motivation for this work is based on several fundamental objectives to advance the integration of artificial intelligence in education, and more specifically in its generative aspect. It introduces a novel AI rapid learning methodology that allows students to face first-hand the challenges and capabilities of these technologies. This hands-on experience serves to stimulate interest in science, technology, engineering and mathematics, the so-called STEM careers, and facilitates a deeper understanding of the practical applications of AI in various fields. In addition, this research explores how our methodology fosters collaboration between students from different schools and educational levels. By participating in collaborative AI projects, it has been observed that students who do not know each other collaborate to solve problems, exchange knowledge, and mutually enhance each other's educational experiences. Therefore, the methodology proposed in this article develops collaborative competency among students \citep{magraner2016educacion}.

A key aspect of this work is to investigate students' perceptions of generative AI following their participation in these experiential learning activities. Understanding high school students' attitudes, concerns, and opinions toward generative AI can provide valuable information about its potential impact on their academic and career paths. For educators, this study aims to demonstrate the pedagogical utility of generative AI \citep{karras2023dreampose,khachatryan2023text2video,gozalo2023chatgpt,karaarslan2024generate,wu2023tune,ali2024constructing,han2023design}. Teachers can leverage this technology to design engaging learning activities that encourage students to explore deeper into other subjects more deeply and creatively, often without them even realizing it. In addition, this research aims to identify the AI applications commonly used by students in their daily lives and to understand the purposes for which they are used. This exploration offers insight into the current integration of AI technology into students' routines and highlights potential areas for further application and improvement. As the exploration progresses, four key \textit{research questions} (RQ) guide this inquiry:

\begin{itemize}
\item \textbf{RQ1.} What emotions do students experience when they think about the evolution of AI and how do they evaluate the ethical implications and responsibilities of using AI tools in their learning?
\item \textbf{RQ2.} Can the use of generative AI foster creativity and innovation in students?
\item \textbf{RQ3.} Do teachers currently encourage the use of AI as a resource to support lectures or assignments?
\item \textbf{RQ4.} Can the use of generative AI help with learning in other disciplines?
\end{itemize}

These questions address some psychological dimensions of AI integration in education ----emotional, creative, and practical, from the learner's perspective. By exploring these questions, a better understanding of learners' perceptions and emotions regarding AI, its developments, and its potential applications is aimed to be provided to educators. In addition, the aim is to deepen the understanding of the role of AI in society and its impact on the professions of the future. Therefore, two main lines of research are pursued in this work. The first one focuses on the methodology of accelerated learning. On the other hand, the analysis aims to understand students' perceptions of AI: how it is used, the possible uses attributed to it, and how they integrate it into their daily lives as a support tool rather than as a definitive solution. Students are aware of the limitations of AI and the need to use it with caution and appropriate knowledge. Furthermore, this study aims to elicit reflections among students and to understand how they see the future of AI and the implications of its evolution.

In the next section \ref{methd}, the methodological part of fast AI learning is explained. Also in this section, the dynamics of the questions are explained. Subsequently, in section \ref{expe}, the experiment, the activity and the results by question type are explained to give a better view of the results. After analyzing the results, the research questions initially posed are addressed in section \ref{RQ_SEC}. Finally, the main conclusions of this study and implications for future work are presented in section \ref{concl}.

\section{AI - student interaction and information retrieval}
\label{methd}
A controlled framework is presented in which secondary school students are exposed to the use of generative AI for subsequent analysis, focusing on its ethical uses and practical applications in their daily lives. This educational program is structured in two main segments to ensure a deep and thoughtful understanding of the subject. In the first part, the basics of AI are addressed, starting with the definition of human intelligence. This section is designed to be formative and dynamic, lasting approximately one hour. During this initial session, strategic questions are posed to students to encourage reflection on the concept of AI and its roots in human intelligence. They are asked what they understand by \textit{intelligence}. Student responses included concepts such as \textit{"ability to solve problems, analyze, reason, innovate, knowledge, logic and critical reasoning, think”}.  After capturing the students' attention and initiating an active dialogue with them, exploration of the concept of artificial intelligence commenced. Despite being a term widely used in today's society, it is observed that it has not been analyzed in depth in its close context. The students' answers to the question of what they understood by artificial intelligence included descriptions such as \textit{“problem-solving machine, imitating human intelligence, improvement of human intelligence, the ability of non-living beings to solve problems, non-human being with the intention of copying us”}. It is remarkable how many students perceive artificial intelligence as a tool intended to enhance human capability.

\begin{figure}
	\centering
		\includegraphics[width=0.6\linewidth]{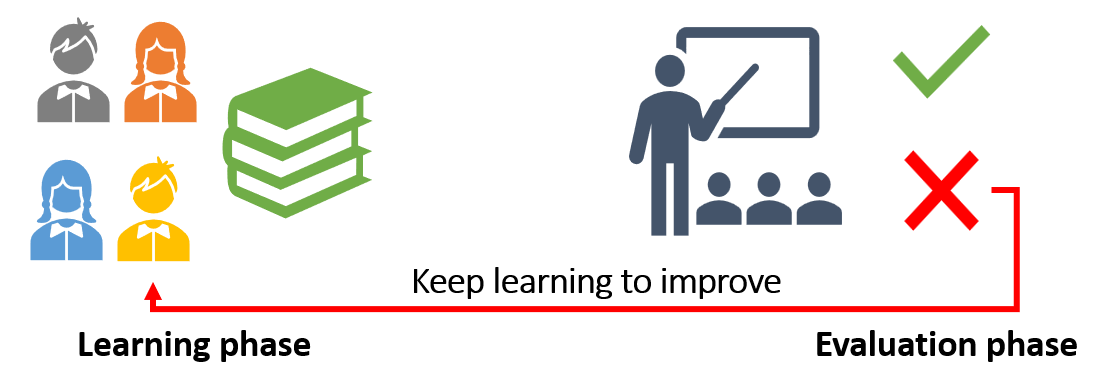}
	\caption{Scheme of explanation of the learning of an algorithm included in the methodology.}
	\label{learn}
\end{figure}

After this introduction, a more detailed analysis of AI is made, focusing on the different learning methods used by AI systems. To make the content more accessible and relevant to students with ages within the interval [15-17], analogies are made between how AI-models learn and their day-to-day lives. A brief introduction to AI is given, explaining its learning methods. To arouse their interest and facilitate their understanding of the topics, it is suggested that when studying for an exam, they act as AI models. They will prepare to learn the subject and to pass the exam by studying the contents found in the study books for further evaluation. If they pass the teacher's assessment, as shown in Figure \ref{learn}, they will have efficiently taken advantage of the training and it will be considered completed. However, if they do not reach the minimum score set by the teacher, it will be necessary to study more for the next exam to re-evaluate the knowledge acquired. 

Explanation of algorithms at a basic level, giving examples from your daily life in which a model such as an autonomous car may be involved. In addition, several end-to-end generative AI applications are presented so that they can feel that it is close and very accessible. One of these applications is \texttt{Doodle-dash}. This application recognizes and interprets the doodles drawn by students (\citep{wolf2019huggingface}). In an educational context it helps students improve their creativity, critical thinking and artistic development.  On the other hand, it is presented the application \texttt{AutoDraw}, which uses recurrent neural networks (RNNs) to recognize doodles drawn and then complete them in an accurate way, being useful to help artistic expression with reduced mobility or \textit{neurodegenerative} disorders such as Alzheimer's and Parkinson's disease \citep{gan2024barriers}, becoming a tool for \textit{democratization of art} \citep{ha2017neural}. It also explains \texttt{Smart Cuisine} as an application to design recipes, following nutrition and culinary science. The goal of this type of program is to learn important skills such as cooking and nutrition, as well as to help a smart use of resources in the kitchen, being a very valuable tool for sustainable cooking, as it can give you recipes with only the ingredients you already have, helping the user to save and intelligently reuse ingredients \citep{kansaksiri2023smart}. This tool is explained to give students an understanding of the great interaction that generative AI can have with other fields. After explaining in a simple way the principles of AI as well as practical approaches that can be used by students as a tool, a short free-response survey with the following two questions is asked to get their opinion:

\begin{itemize}
\item What tools have you used and what for? (See No. 24 in Table \ref{free_respo})
\item In what situations in your day-to-day school life could you use AI? (See No. 25 in Table \ref{free_respo})
\end{itemize}

\begin{figure}
	\centering
		\includegraphics[width=0.9\linewidth]{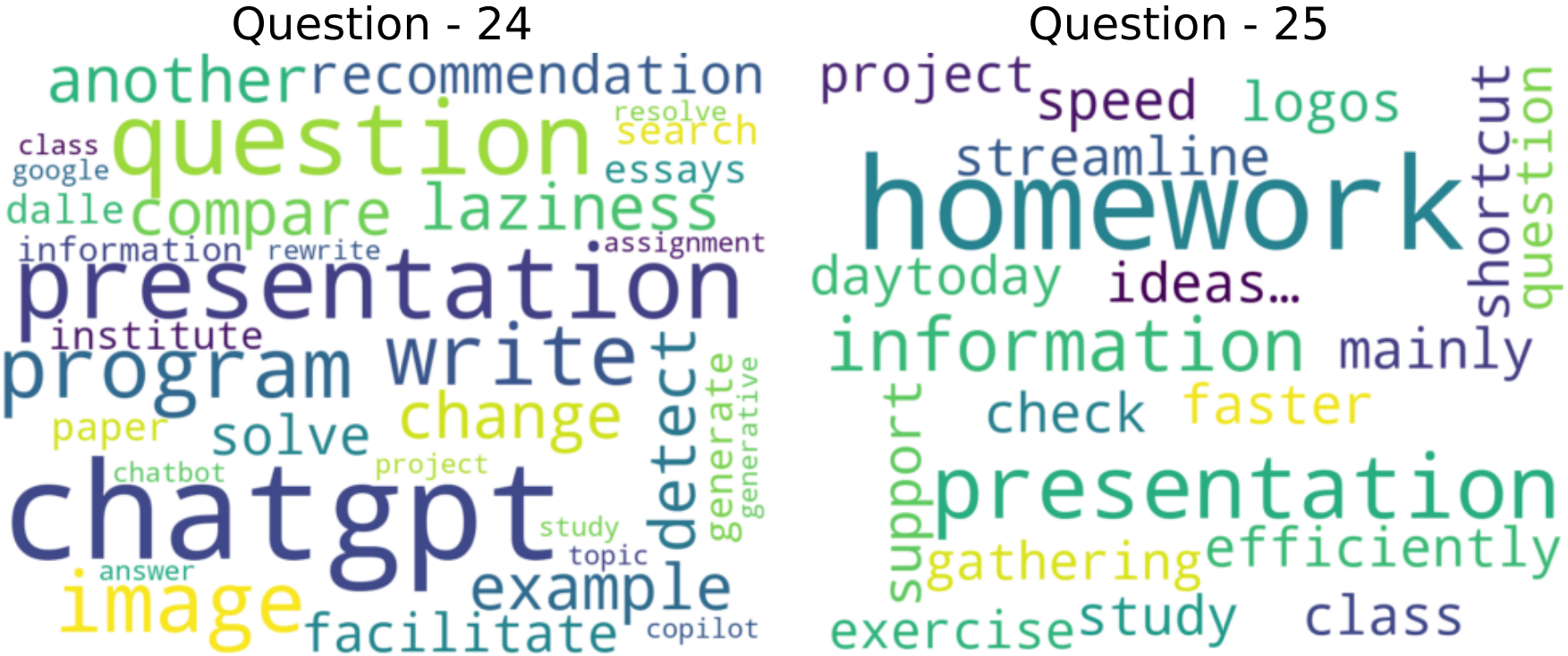}
	\caption{Wordcloud by frequency obtained from the free-response answers made during the activity.}
	\label{WC_DURING_AC}
\end{figure}

From these free-response questions, different answers of forms of use by each student emerged, although they had several in common. In Figure \ref{WC_DURING_AC} it can be seen the \textit{Wordcloud} with the most frequent words in the answers for both questions 24 and 25 respectively. As can be seen, the most used tools include \texttt{ChatGPT} for text generation, \texttt{Copilot} \citep{perez2020copilot} and \texttt{Dall-e} as an image generator based on transformers \citep{vaswani2017attention}. Other tools also include \texttt{Chatbots} and several programs for creating presentations and obtaining recommendations. When asked what they used them for, they answered that they used them to solve problems faster, compare and check information and rewrite results obtained by other AI tools to humanize the results. Interestingly, one finding is that some students use tools such as \texttt{ChatGPT} in times of boredom to talk, indicating some emotional interaction with the platforms \citep{bartsch2010use,tan2008entertainment}. On the other hand, students were asked to think about possible uses they could make of AI in their daily lives. In Figure \ref{WC_DURING_AC}, it is found that most of them would use it for their daily homework, although there are other interesting uses such as checking their work, generating logos, getting information faster, getting inspired with new ideas, finding shortcuts, etc., which allows students to improve and achieve results more efficiently. Some students indicated that they used these applications to generate questions to help them study, a clear example of the use of AI as a tool to support the study of other subjects.

This was followed by an interactive activity in which all students actively participated. They used artificial intelligence tools to create videos from free prompts, using notebooks that employ stable diffusion models \citep{Deforum_v071}. To assess the students' perception of this experience, a survey was conducted that addressed several dimensions, both sociological and psychological. The objective was to understand how the integration of artificial intelligence in the classroom can influence students' attitudes, learning and behavior.

\begin{table}[h]
    \centering
    \caption{Sentences and questions proposed in the questionnaire with responses rated according to the Likert scale \citep{schrum2020four} according to their satisfaction.}
    \label{likert_}
    \begin{tabular}{|c|p{0.7\linewidth}|}
        \hline
        \textbf{ID} & \textbf{Question - Likert scale (1: Totally disagree - 3:Neutral - 5: Totally agree)} \\
        \hline
        1 & \textit{The AI learning methodology has allowed me to actively participate in the class.} \\
        \hline
        2 & \textit{The activity has facilitated interaction and collaborative work with my classmates.} \\
        \hline
        3 & \textit{I have learned about the different functionalities and practical uses of AI as a tool.} \\
        \hline
        4 & \textit{During this learning methodology, I have learned how collaboration between different technologies can generate innovative results.} \\
        \hline
        5 & \textit{I consider that I can use the tools learned for the performance of classroom subjects.} \\
        \hline
        6 & \textit{I consider that the interaction with AI tools in class has been interesting and enriching for my learning.} \\
        \hline
        7 & \textit{I am more interested in exploring AI-related topics after participating in this learning methodology.} \\
        \hline
        8 & \textit{I feel more confident using AI technologies after participating in this learning methodology.} \\
        \hline
        9 & \textit{I have felt frustration while participating in this AI learning methodology.} \\
        \hline
        10 & \textit{I am more motivated to learn about AI-related topics outside the school environment thanks to this learning methodology.} \\
        \hline
        11 & \textit{This AI learning methodology has fostered my creativity and innovative thinking.} \\
        \hline
        12 & \textit{The AI learning methodology has sparked my interest in exploring technology-related careers in the future.} \\
        \hline
        13 & \textit{Overall, I am satisfied with the AI learning methodology and its applications in this class.} \\
        \hline
        14 & \textit{Would you recommend this AI learning methodology to other high school students or to your high school teachers?} \\
        \hline  
    \end{tabular}
\end{table}

\begin{table}[h]
    \centering
    \caption{Questions asked about your experience with binary response}
    \label{tab_yn}
    \begin{tabular}{|c|p{0.7\linewidth}|}
        \hline
        \textbf{ID} & \textbf{Question (YES/NO)} \\
        \hline
        15 & \textit{Would you leave your usual activities/duties in the hands of an AI system?} \\
        \hline
        16 & \textit{Do your regular teachers encourage you to use any AI applications as support material?} \\
        \hline
        17 & \textit{Have you received any training on how to use AI tools ethically and responsibly in your learning?} \\
        \hline
    \end{tabular}
\end{table}

\begin{table}[h]
    \centering
    \caption{Free-response questions.}
    \label{free_respo}
    \begin{tabular}{|c|p{0.7\linewidth}|}
        \hline
        \textbf{ID} & \textbf{Question (Free response)} \\
        \hline
       18 & \textit{If in question No. 9 you answered with a score of 3-5, briefly explain why.} \\
        \hline
        19 & \textit{How has the AI learning methodology inspired you to be more creative? Explain briefly.} \\
        \hline
        20 & \textit{Briefly explain why you would recommend this methodology.} \\
        \hline
        21 & \textit{What aspect of the activity do you consider most outstanding?} \\
        \hline
        22 & \textit{Explain briefly why you would entrust your activities to AI.} \\
        \hline
        23 & \textit{Briefly describe your emotions when thinking about the evolution of AI.} \\
        \hline
        24 & \textit{What tools have you used and what for?} \\
        \hline
        25 & \textit{In what situations of your daily school life would you use it?} \\
        \hline
    \end{tabular}
\end{table}

The survey questions are structured in six distinct analytical dimensions, each focusing on different aspects of the use of artificial intelligence in education. This multidimensional approach facilitates a comprehensive analysis that examines both the impact at the individual level, from a psychological perspective, and the impact on the group and institutional context, from a sociological perspective.

\begin{enumerate}
    \item \textbf{Learning and Participation Experience (LPE)}: This dimension encompasses the student's overall experience in the learning process and their active participation in the class. The questions related to this dimension are 1, 2, 3, 5, 13, which appear in Tables \ref{likert_}, \ref{tab_yn} and \ref{free_respo} respectively. All the proposed questions focus on the student's experience during the learning process and how the AI methodology influenced their participation and understanding of new technologies \citep{quay2003experience}.

    \item \textbf{Impact and Motivation (IM)}: It addresses the impact of the methodology on student motivation and interest in AI-related topics. The questions that assess this dimension are 6, 7, 10, 14, 24 from the same tables. The selected questions provide information about how the methodology affects student motivation to learn and explore more about AI and its uses \citep{ferriz2020gamification}.

    \item \textbf{Emotional Experience (EE)}: It refers to the emotions and feelings experienced by the students during the learning session. The questions 8, 9, 18, 23, seek to understand the learner's emotions concerning frustration and perception about AI evolution \citep{henry2003emotional}.

    \item \textbf{Creativity and Innovative Thinking (CPI)}: It evaluates how the AI methodology has fostered student creativity and innovative thinking. Questions 4, 11, 19, focus on how the methodology has influenced the student's ability to think creatively and innovatively \citep{seechaliao2017instructional}.

    \item \textbf{Future Perspective and Orientation (FOP)}: Focused on the student's future vision and orientation toward future STEM careers and applications. The questions 12, 16, 20, 25 of the questionary assess the influence of the methodology on the student's orientation toward future \citep{sulistiobudi2023employability}.

    \item \textbf{Ethics and Responsibility (ER)}: It addresses student awareness of ethics and responsibility in the use of AI. Specifically, questions 15, 17 and 22 are designed to gauge students' perception of their ethical development and responsibility in AI usage, as well as their readiness to delegate tasks to AI systems \citep{mulang2023exploring}. 
\end{enumerate}

In the following section, it is found a detailed explanation of the activity as well as the results obtained from the surveys and the students' perception to draw statistical conclusions.

\section{Experiment and results}
\label{expe}

The methodology proposed in this study is pilot-tested with ten high school students from different schools in Madrid, Spain. In the present study, the sample composition includes 90\% male participants and 10\% female participants. Note that, although this disparity in gender representation exists, the scope and main objectives of the analysis are not oriented to explore gender differences. Consequently, the observed imbalance in gender ratio is not considered a determining factor for the results and conclusions derived from this research. 

\begin{figure}
	\centering
 \includegraphics[width=0.7\linewidth]{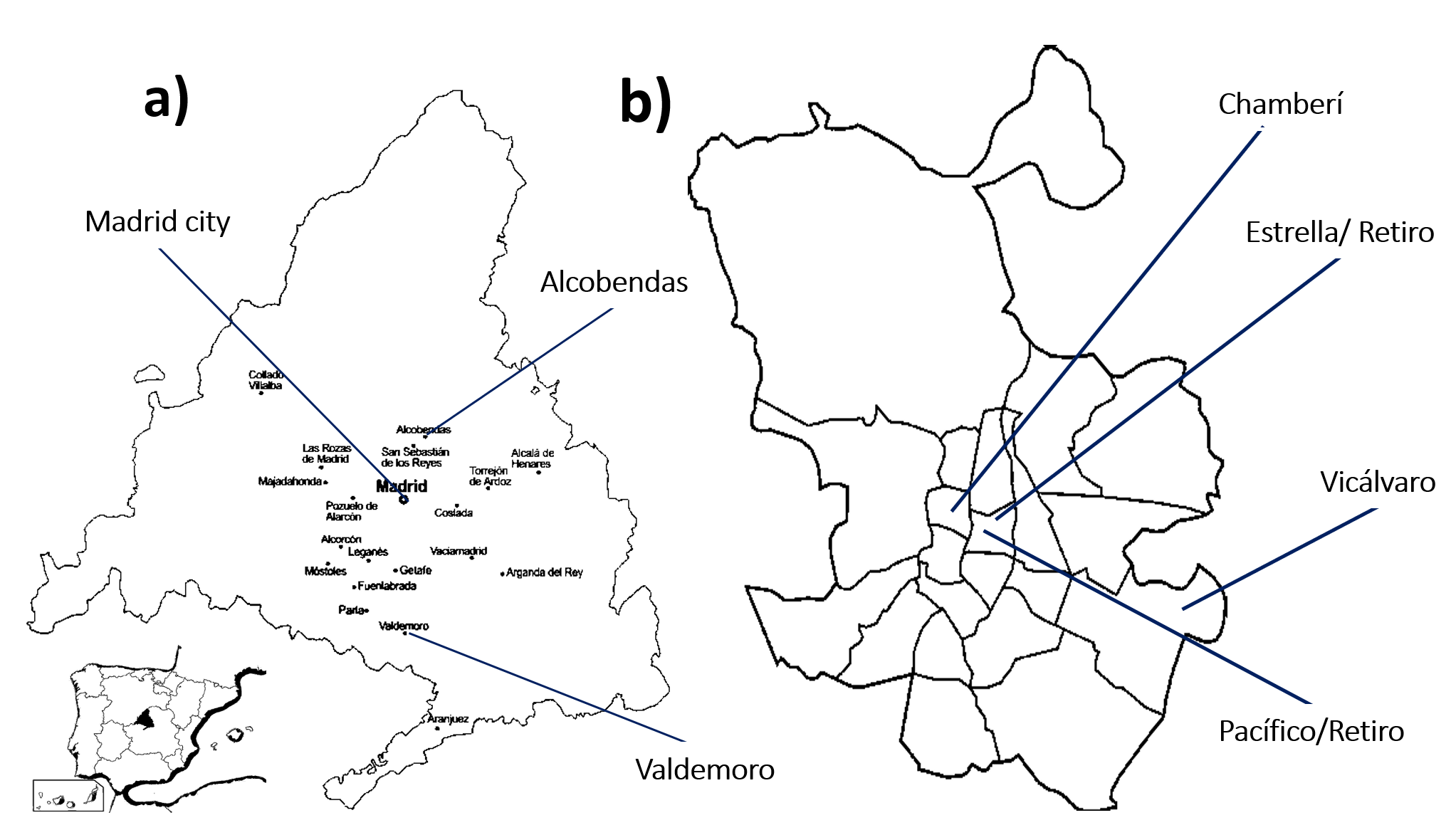}
	\caption{a) Map of the community of Madrid with municipalities outside the city of Madrid where there are participating public centres. b) Districts of Madrid city with participating public centres.}
	\label{mapas}
\end{figure}

Although our study sample comprises a limited number of students, they have been selected from the \textit{“4ºESO + Empresa”} program of the community of Madrid, which provides strategic representativeness within the educational context of public schools in the Community of Madrid \citep{4eso}. This program is designed to offer secondary school students an approach to the world of work through training placements in companies, allowing students to explore career opportunities and learn first-hand about working in companies in the sector they are most interested in. Note that in the context of this program students freely choose based on their interests the company they would like to assit. The sample studied includes students from different districts and municipalities of the community of Madrid as can be seen in Figure \ref{mapas}, involving public centres from urban areas to the north and south of the capital of Spain as shown in Figure \ref{mapas}a, as well as from different centres in different districts of the city of Madrid as can be seen in Figure \ref{mapas}b, which ensures that both urban and suburban areas are represented. 

In the sample of students, five of them correspond to the same school, while the other five are from different schools. Since the students come from different schools and different municipalities, they do not know each other. This suggests that the level of interaction in the activities is lower among students who do not know each other than among those who share a centre, which allows the present methodology to validate the capacity for interaction and collaboration among students \citep{johnson1981student}. The activity has a maximum duration of three hours, which is a challenge given the lack of knowledge of the techniques to be explained. 

After establishing the context and the challenges inherent to the activity, four basic themes are proposed for the generation of the videos. These themes are proposed in order to homogenize the results and to be able to compare the approaches written in the prompts, the so-called \textit{prompt engineering} \citep{white2023prompt}, and to favor interaction and collaboration among the students:

\begin{enumerate}
 \item Evolution of civilizations.
 \item Fractals.
 \item Cities as different artist styles.
 \item Evolution of clothing.
\end{enumerate}

\begin{figure*}
	\centering
 \includegraphics[width=0.8\linewidth]{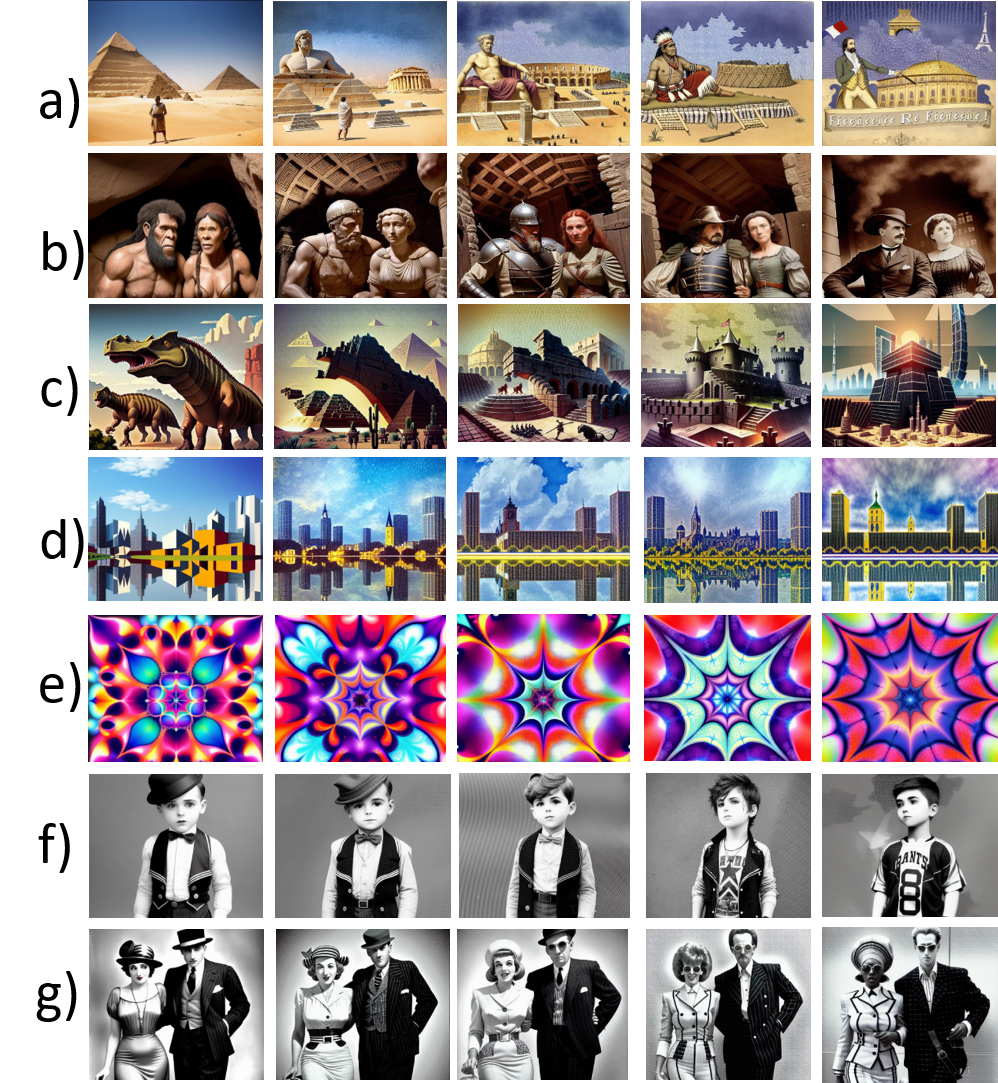}
	\caption{Videos produced by students' free prompts. a) A man in egypt with pyraminds in the background, a man in the ancient greece, a roman emperor during the fall of roman empire, a native american being colonized, a french republican during french revolution. b)Primitive man ans women in a cavern, man and women of the roman empire, man and women archer in a castle, mustketeer and a woman musketeer in the medium age, man and women in the wall street of new york. c) a prehistorical landscape with a bunch of dinosaurs trending on artstation, four pyramids in the dessert with a sunny sky, the roman coloseum with a gradiator fighing against a lion, an historical castle with with a trench under a storm, a city of Dubai in a sunny day with their skyscrapers. d) a beautiful city with cubism style,  a beautiful city with Van Gogh style,  a beautiful city with Da Vinci style, a beautiful city with Velazquez style,  a beautiful city with Picasso style. e) Fractals, vivid colors. f) a beautiful boy dressed like 1920s style, a beautiful boy dressed like 1940s style, a beautiful boy dressed like 1960s style, a beautiful boy dressed like a rock style, a beautiful boy dressed with sport clothes. g) Man and a woman dressed like in the 1920s, Man and a woman dressed like in the 1940s, Man and a woman dressed like in the 1960s, Man and a woman dressed like in the 1980s, Man and a woman dressed like in the 2000s.}
	\label{videos}
\end{figure*}

All groups made 100 frames so that the changes in the prompts could be seen, but they were limited by the time of the session. Two hours were left for the activity, during that time the students showed high participation and collaboration, improving the prompts by sharing various aspects or bad results as described. The students checked how parameters such as zoom or angle could affect the final results. All of them used 2D animations and a frequency of 4 seconds of video per 50 frames generated. Eight-second videos were generated at a rate of 12 frames per second, i.e. approximately 100 frames per video in total. Each video consisted of five successive prompts, as shown in Figure \ref{videos}, of twenty frames each. While making the videos, the students realized that sometimes they did not get the results they expected from the AI, so they used \textit{negative prompts} with the same frame rate to better guide the generative model and get more accurate results from, in this case, \textit{deforum stable diffusion} \citep{Deforum_v071} which is an open source software and provides significantly better results than other non-diffusion based models such as generative adversarial networks (GANs) \citep{dhariwal2021diffusion}.

Each working group was composed by two students. Within the global themes, students are given the freedom to write the prompts, which encourages creativity and innovation of the outputs. Note that although there are similar elements because the themes were bounded, there are many elements that differ as can be seen in Figures \ref{videos}a,b,c. During this activity, some students detected the presence of bias in artificial intelligence, which led to an enriching ethical debate. Through a historical example observed in the first frames of Figure \ref{videos}g, it was discussed how in the 1920s, the AI did not recognize the possibility of black people belonging to higher socioeconomic classes in American society. However, in stills from the 2000s, a significant change is observed in the same figure, where AI is already able to recognize and anticipate without difficulty the existence of families composed of black women and white men in the upper classes. This progress reflects how AI has adapted to the evolution of racial integration in contemporary society. Although there is still work to be done, it is clear that AI systems are moving toward more ethical and less discriminatory practices, in line with social evolution. Students understood that the data on which these AI models are trained are generated by humans, which highlights the importance of being meticulous and ethical in their selection and handling. Understanding these aspects is not only a fundamental part of learning, but also key to fostering inclusive, equitable and quality education, in line with the United Nations \textit{Sustainable Development Goal} (SDG) number 4: \textit{quality education}, \citep{ferguson2020sdg, sdg2019sustainable}.

Next, the answers to the questionnaire made by the students at the end of the activity will be analyzed in detail. This survey has questions of different categories, so they will be analyzed separately, although an overall conclusion will be given at the end of the activity.

\subsection{Binary questions analysis}

The proposed binary questions aim to understand the level of training of students from different schools regarding the ethical uses of artificial intelligence, as well as to investigate whether teachers encourage its integration into their teaching methods.

\begin{figure*}
	\centering
	\includegraphics[width=0.8\linewidth]{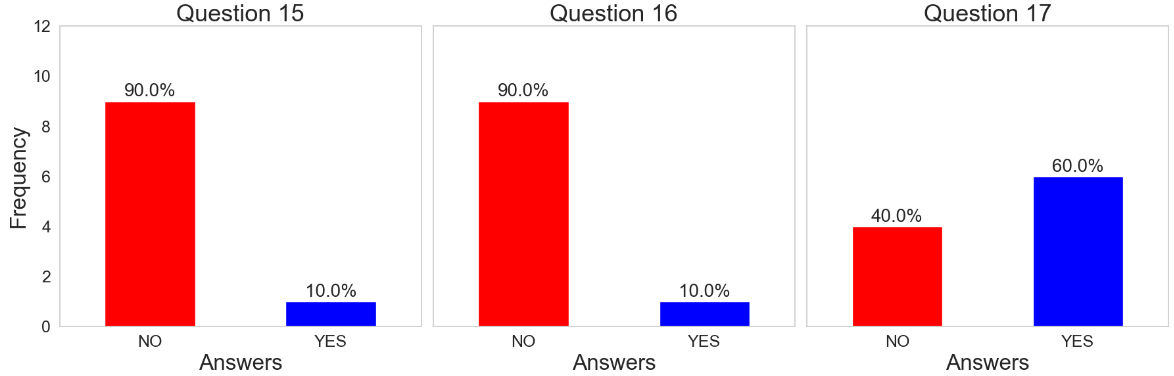}
	\caption{Binary responses.}
	\label{bina}
\end{figure*}

In question 17, 60\% of students reported that they have received training on the ethical use of AI as can be seen in Figure \ref{bina}. However, a striking contrast is evident from the responses to question 16, where only 10\% of students reported that their teachers actively promote or facilitate the use of AI in their academic pursuits. This disparity highlights a potential area for professional development among educators. There is a notable opportunity to enhance teacher training, focusing on the integration and pedagogical applications of AI across various disciplines. Additionally, it was observed that students frequently turned to online resources to gather more detailed information on subjects such as history or art to better tailor their AI-driven prompts. This behavior suggests an enhancement in students' research and academic study skills facilitated by the utilization of AI technologies. The proactive use of AI in educational settings appears to contribute positively to the interdisciplinary research capabilities of students \citep{kamposiori2022embedding}.
In question 15, a significant majority (90\%) of the students responded negatively to the idea of entrusting their activities to artificial intelligence. The subsequent question, question 22, sought to elucidate the reasons behind their reluctance. Analysis of the responses reveals frequent use of the words \textit{mistake} and \textit{help}. Some students' explanations are:

\begin{itemize}
    \item \textit{"You never know if the AI can make a mistake. After all, it is a replica of us, and we make mistakes. They might make fewer mistakes, but they can still make them."}
    \item \textit{"Even though it has performed well, there have been instances when it did not think the same way I did."}
    \item \textit{"Not entirely reliable."}
    \item \textit{"May be flawed."}
     \item \textit{"To help me but not to do everything"}
\end{itemize}

Only the 10\% of the students indicated that they would leave their activities in the hands of AI focused on AI's capabilities to help human beings, \textit{"Because it seems to me that it is very well qualified to help human beings."} Overall, these comments reflect a general concern about the potential for errors and a lack of complete reliability in AI systems, influencing the students' trust in such technologies for managing their activities.

\subsection{Likert scale questions analysis}

\begin{figure*}
	\centering
		\includegraphics[width=\linewidth]{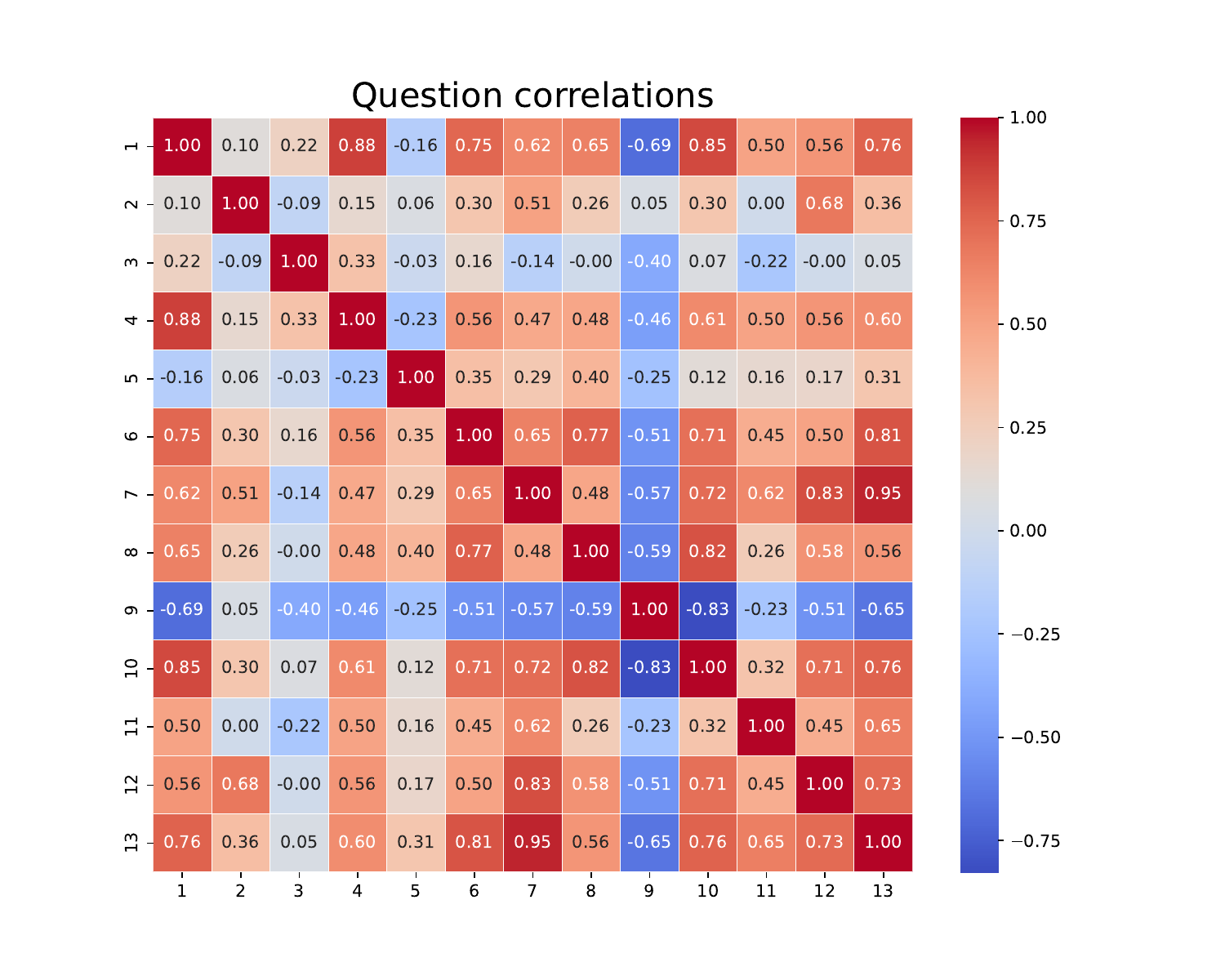}
	\caption{Correlation matrix among the Likert scale questions.}
	\label{corr}
\end{figure*}

Once the students' responses to the Likert scale questions were collected, it is proceeded to their analysis using descriptive statistics. The results obtained are shown in Table \ref{likert_stats_updated}. Likewise, Figure \ref{corr} shows the study of correlations between the questions in Table \ref{likert_} belonging to the dimensions LPE, IM, CPI, EE and FOP to better understand how they relate to each other and to understand student behaviors. In this analysis, special attention is paid to correlations whose values, positive or negative, exceed the threshold of 0.75, since this indicates a strong association between the questions analyzed. Note that question 14 is not included in the correlation matrix since it does not provide useful information on the relationships between the variables, since as can be seen in Table \ref{likert_stats_updated}, $\sigma_{14}$ =0, there being no dispersion in the sample responses.

Analysis of the survey responses reveals that active class participation is closely linked to several positive outcomes in student learning and motivation concerning AI. First, there is a high correlation (0.88) between active participation (question 1 from Table \ref{likert_}) and learning how different technologies can collaborate to generate innovative results (question 4 from Table \ref{likert_}). The proposed methodology encourages students' active participation ($\mu_1$ = 4.8), thus enhancing AI learning through collaboration between technologies ($\mu_4$ = 4.4). Moreover, this active participation that promotes this methodology favours significantly the students' motivation to explore AI-related topics outside the school setting ($\mu_{10}$ = 4.1). Finally, active participation also has a significantly high correlation with the interaction of AI tools (question 6 from Table \ref{likert_}) as an interesting experience within the classroom ($\mu_6$ = 4.5) and with overall satisfaction with the proposed methodology (question 13 from Table \ref{likert_}, $\mu_{13}$ = 4.7).

\begin{table}[h]
\caption{Descriptive statistics of Likert scale evaluated questions, where three is the Neutral status and $n$ =10.}
\centering
\begin{tabular}{ccccc}
\hline
ID Question & $\mu$ & Mode & $\sigma$ & Wilcoxon \\ \hline
1        & 4.8   & 5    & 0.421    & 0.00195  \\
2        & 4.6   & 5    & 0.516    & 0.00195  \\
3        & 4.7   & 5    & 0.483    & 0.00195  \\
4        & 4.4   & 5    & 0.843    & 0.00829  \\
5        & 4.3   & 4    & 0.674    & 0.00617  \\
6        & 4.5   & 5    & 0.707    & 0.00575  \\
7        & 4.5   & 5    & 0.849    & 0.00665  \\
8        & 4.0   & 4    & 0.816    & 0.01518  \\
9        & 1.8   & 1    & 0.918    & 0.01387  \\
10       & 4.1   & 4    & 0.994    & 0.01762  \\
11       & 4.5   & 4    & 0.527    & 0.00195  \\
12       & 4.0   & 4    & 0.942    & 0.01969  \\
13       & 4.7   & 5    & 0.483    & 0.00195  \\
14       & 5.0   & 5    & 0.000    & -       \\\hline
\end{tabular}
\label{likert_stats_updated}
\end{table}

The high correlation (0.83) between question 7 and question 12 from Table \ref{likert_} indicates that increased student interest in exploring AI-related topics after participating in the learning methodology is strongly associated with increased interest in exploring technology-related careers in the future. This suggests that the classroom experience is not only sparking increased interest in the AI subject itself ($\mu_7$ = 4.5), but is also inspiring students to consider career options within the technology field ($\mu_{12}$ = 4.0). This finding is significant as it indicates the potential positive impact that AI learning methodology can have in shaping students' future career paths. Apart from that, the high correlation (0.95) between students' interest in AI-related topics (question 7 from Table \ref{likert_}) and satisfaction with the learning methodology used (question 13 from Table \ref{likert_}) suggests that students' satisfaction favors their curiosity to learn more about AI. As student satisfaction is very high ($\mu_{13}$ = 4.7), it shows that this methodology favors students' interest in AI topics.

On the other hand, a notable observation emerges when examining the high correlation between questions 6 and 8, indicating that an experience perceived as interesting and enriching with AI tools in class is strongly associated with an increase in students' confidence in using these technologies. In the proposed methodology, students perceive that interaction with AI tools is positive and beneficial to their learning ($\mu_6$ = 4.5), so they also feel more confident and competent when using these technologies ($\mu_8$ = 4.0). Furthermore, the correlation of 0.81 between question 6 and question 13 reveals that a positive and enriching experience with the interaction of AI tools in class is closely linked to a higher overall satisfaction with the AI learning methodology. In this methodology, as students rate their experience with AI tools highly, they also have a very positive and satisfying perception ($\mu_{13}$ = 4.7) of how the subject is taught in the classroom.

Question 10 has a notable positive correlation with question 8 (0.82) and with question 13 (0.76). This means that increased student motivation to learn about AI-related topics outside the school environment, as reflected in question 10, is highly correlated with both a greater feeling of safety in using AI technologies, as indicated by question 8, and greater overall satisfaction with the AI learning methodology and its applications in the classroom, as reflected in question 13. These high correlations suggest that when students show greater interest and motivation to explore AI outside the classroom due to the learning methodology used, they also tend to feel more confident in using these technologies and are more satisfied with the way they are taught the subject matter in the classroom. This finding is significant because it suggests that the AI learning methodology is not only succeeding in increasing students' motivation to learn about the subject ($\mu_{10}$ = 4.1), but it is also positively impacting their confidence ($\mu_8$ = 4.0) and satisfaction ($\mu_{13}$ = 4.7) regarding how the subject is taught in the classroom. On the other hand, question 10 has a high negative correlation with question 9. This indicates that as students' perception of frustration decreases during their participation in the AI learning methodology (question 9 from Table \ref{likert_}), their motivation to learn about AI-related topics outside the school environment increases (question 10 from Table \ref{likert_}). This methodology favors that frustration is very low ($\mu_9$ = 1.8), and, therefore, that motivation to inquire about AI is high ($\mu_{10}$ = 4.1).
Finally, note that the value three on the Likert scale reflects a neutral opinion. The students never responded to this option when answering the questionnaire, as is often the case in many studies, in which participants simply chose the intermediate option to not participate or to respond quickly.

This analysis of correlations and dimensions relies on the Wilcoxon test \citep{wiedermann2013robustness} to evaluate the significance of responses with respect to the neutral value on Likert scales. This non-parametric test is suitable for ordinal data and does not require assumptions of normality, being useful even in small samples. The results of the Wilcoxon test are presented in Table \ref{likert_stats_updated}. The findings indicate very low \textit{p-values} for all questions, suggesting significant differences between the responses and the neutral value of the Likert scale. This discrepancy highlights a trend where responses tend to be polarized rather than neutral, implying that students have displayed attitudes that are not passive or neutral, but rather demonstrate satisfaction with the evaluated methodology. It is important to note that question 14 lacks a p-value due to zero variability in the responses ($\sigma_{14} = 0$), meaning all responses were identical and therefore not suitable for significance testing. This uniformity in responses eliminates the possibility of conducting a meaningful statistical analysis for this particular question, but in itself as all responses were 5/5 on the Likert scale it can be concluded that all students recommend this methodology.

\subsection{Free responses analysis}

\begin{figure*}
	\centering
	\includegraphics[width=0.9\linewidth]{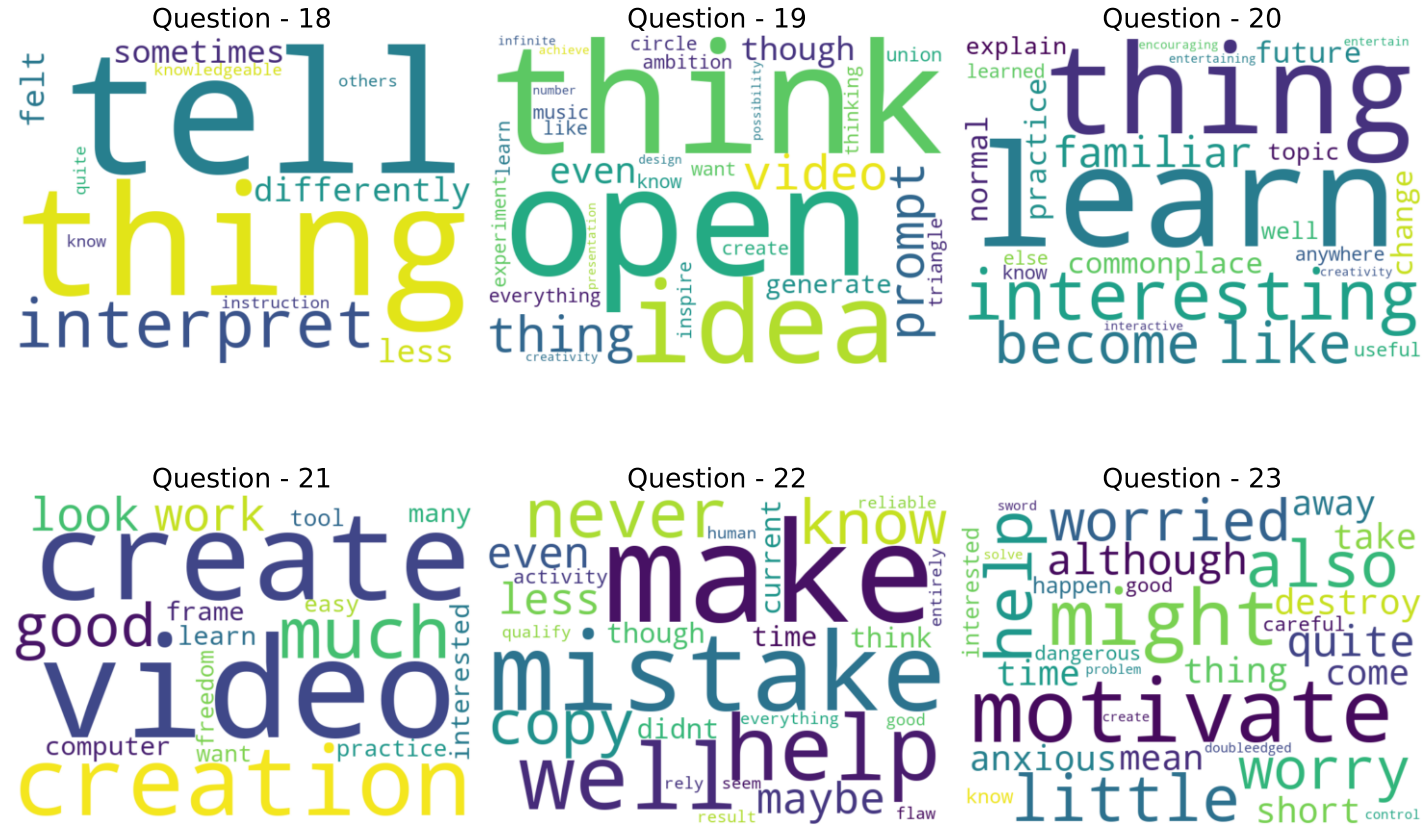}
	\caption{Wordcloud by frequency obtained from the free-response answers made at the end of the activity.}
	\label{WC_2}
\end{figure*}

Next, a deeper analysis of the free responses provided by the students in the questionnaire will be conducted. The main objective is to gain a deeper understanding of their opinions, perceptions and experiences related to the methodology and the proposed activities. The specific questions to be addressed in this analysis are detailed in Table \ref{free_respo}. These questions were designed to capture the direct impressions of the students, allowing them to freely express their thoughts and feelings about the activity and their experience with artificial intelligence. To facilitate understanding and visualization of the most recurring themes and words in the students' responses, word-clouds are created as shown in Figure \ref{WC_2}. These word-clouds highlight the most frequently used words, providing a visual representation that helps to quickly identify the main themes and predominant opinions among students. Analysis of the free responses and word-clouds will provide us with a more complete and nuanced view of student perceptions. They will allow us to identify patterns, trends, and areas of interest or concern, which in turn will help us better understand how students perceive and value the methodology and their interaction with artificial intelligence.

When analyzing question 9, which addresses the frustration experienced during the activity, it is observed that the mean is 1.8, with the mode being 1. This suggests that most students felt actively engaged in the activity and did not experience a great deal of frustration. On the other hand, question 18 focused on understanding the reasons behind any frustration or confusion students may have felt. This question was only answered if students expressed feeling frustrated or lost during the activity. Most indicated that their frustration was due to the AI interpreting the prompts in a different way than they had thought. A representative response was: \textit{"Because I told the AI one thing and it interprets it differently from me."}. This response suggests that the trainee experiences a sense of doubt regarding the AI, questioning whether the AI is always right. This discrepancy in interpretation may influence the students's perception of the AI's accuracy and reliability, or lead to a decrease in the trainee's confidence in his or her abilities, perceiving that his or her understanding or presentation of information does not match the level of accuracy or objectivity possessed by the AI \citep{singh2023review,famaye2023ban}.

These results point to a key understanding on the part of the learners: what they understand and communicate does not necessarily match what the AI models understand. This is also due to the lack of interpretability of models. This discrepancy in interpretation between learners and AI highlights the importance of being careful and clear in explanations and communications with AI. Students need to understand that machines interpret information differently than humans and that it is necessary to be precise and detailed in instructions and explanations. This learning can be a valuable lesson for students, not only in the context of the specific activity but also in their future interaction with AI technology.

Question 19 focuses on how the applied methodology has contributed to enhancing students' creativity. The responses show a positive response and reveal several dimensions in which the methodology has had an impact on students' creativity. First, students indicate that the methodology has allowed them to think more openly and with a greater number of ideas \textit{"Thinking in a more open way and with more ideas"}, suggesting that it fosters divergent thinking, in which multiple perspectives are explored, which is fundamental to creativity. In addition, students mention the \textit{“infinite amount of possibilities”} that can be achieved with the use of AI. It gives them a wide range of options to explore and experiment with, enhancing their creativity by not limiting them to a single solution or approach. An interesting aspect is that, although the AI generates the content, learners feel responsible for thinking about what they want the AI to accomplish \textit{"Even though the AI generates everything for you, you are the one who thinks about what you want the AI to do."}. This perception reflects a human-machine collaboration, in which the participant guides and personalizes content production, which could contribute to their sense of authorship and creativity. In addition, the methodology has inspired learners to investigate previously unknown topics to improve their prompts, seeking information from other sources and feeling ambitious to explore them. This suggests that the methodology not only enhances creativity but also promotes self-directed learning and intellectual curiosity, underscoring the positive impact of the methodology on student learning. Interestingly, the use of these technologies makes students have to do more research on the topics to be generated, as prompts must be specified precisely. If translated to any other subject, students must research the topics precisely, which leads them to study without realizing it. In addition, a greater depth of learning and ability to synthesize content is encouraged for prompt engineering, which promotes knowledge acquisition.

On the other hand, question 20 seeks to understand the reasons why students would recommend the applied educational methodology. The answers provided reflect a variety of perceptions about the methodology and its benefits, both educational and personal. Many students emphasize that the methodology allows them to become familiar with concepts and tools that will be common in the future. This suggests that students recognize the value of acquiring skills and knowledge relevant to the current and future world. The opportunity to use generative AI to explain different real topics innovatively is seen as a point in favor \textit{"Because it is a practice that changes from the normal and you can use it to explain a topic as well"}, as it allows to approach the contents more dynamically and attractively, valuing positively that the proposed teaching methodology differs from traditional educational practices. In addition, the responses highlight that the methodology is interesting, fun and entertaining. It is important since an attractive learning environment can increase students' motivation and engagement, thus facilitating the learning process. Finally, the interactivity of the methodology is highlighted as a positive aspect, finding it interesting, entertaining and interactive, which enhances the learning experience by actively involving students in the process.

Question 21 seeks to identify the aspects that students consider most notable in the activity related to the implemented methodology. Students highlight the ability of artificial intelligence to perform tasks and the quality of the results obtained shown in Figure \ref{videos}. The perception of \textit{“how much AI can do”} suggests that students are impressed and surprised by the capabilities of AI, which may foster further interest and curiosity towards this emerging technology. Students also value the freedom offered by the methodology to be able to create and express themselves \textit{“freedom to create whatever we want”}, as shown in Figure \ref{videos}a,b,c, where with the same proposed base theme, the different groups achieved completely different results in terms of eras and civilizations. This is an aspect that highlights the autonomy and personalization that the activity provides, allowing students to develop their creativity and uniquely express their ideas. As shown in the Wordcloud of Figure \ref{WC_2}, the word \textit{videos} and \textit{creation} turn out to be frequent in the responses, denoting that being able to obtain a visual result of the activity boosts motivation.

In question 23 students are asked about their opinion on the advancement of AI in society. This question seeks to know their perception and at the same time to make them reflect on its application and ethical limitations. On the one hand, many students show motivation and enthusiasm towards the potential of AI, expressing interest in the advanced capabilities of AI and how it can lead society towards significant advances. Phrases such as \textit{“Motivated because it is increasingly doing more things, and that means more help”} or \textit{“motivated”} illustrate this optimistic perception, in which students recognize the transformative potential of AI. In parallel, there is a palpable concern among students about the potential risks and challenges associated with AI. There is a sense of unease and caution about the limits and possible negative consequences. Responses such as \textit{“I am a bit worried that it will get out of control”}, \textit{“I am interested but it can be dangerous”}, \textit{“It is a double-edged sword and I worry that it will create more problems than it solves”}, reflect this perception of AI as a tool, with potential for both good and evil depending on how it is used \textit{“Well, but you have to be careful in who uses it or how they use it”}. As can be seen in Figure \ref{WC_2}, the most frequent words in the responses are \textit{motivated} and aspects of concern such as \textit{worry, worried} or \textit{destroy}.

After conducting a thorough analysis of the numerical and free responses provided in our survey, the initial questions posed at the beginning of this study can now be addressed. These questions will be answered using the findings and data obtained, enabling us to provide an informed and detailed assessment of the issues investigated.

\section{Research questions discussion}
\label{RQ_SEC}
In this section, the findings obtained are discussed to answer the initial research questions, with a focus on students' perceptions of the integration and impact of artificial intelligence in education. The answers provided by students offer us valuable insight into their emotions, attitudes and experiences related to the evolution of AI, its potential to foster creativity, the attitude of teachers towards its use, and its ability to enrich learning in different disciplines. Through this analysis, a better understanding of how students perceive and value AI as an educational tool is sought, identifying opportunities and challenges in its integration in the educational setting.

\subsection{RQ1. What emotions do students experience when they think about the evolution of AI and how do they evaluate the ethical implications and responsibilities of using AI tools in their learning?}

When considering the evolution of artificial intelligence, students experience a range of diversified emotions. On the one hand, many are excited and motivated by the transformative potential of AI, seeing opportunities for innovation and advancement in a variety of fields, including education. This optimistic outlook reflects an openness to the possibilities that AI can offer in terms of improving teaching, facilitating learning and personalizing education. On the other hand, not all students share a completely positive view. There are feelings of caution and concern regarding the potential risks and ethical challenges that AI might pose. These sentiments are reflective of a growing awareness among students of the ethical responsibilities involved in the use of AI tools. This ethical awareness manifests itself in a self-directed learning curve, where students show an interest and commitment to understanding the ethical implications and responsibilities associated with the use of AI in their learning. In this sense, students are not only interested in harnessing the capabilities of AI but also show a genuine concern for using it ethically and responsibly in their educational process.

\subsection{RQ2. Can the use of generative AI foster creativity and innovation in students?}
The implementation of generative artificial intelligence positively influences students' creativity and innovation. This technology provides the students the freedom to create and experiment, which encourages the development of their imagination and innovative thinking. The hands-on approach with AI-based challenges motivates students to explore new methodologies and topics, which contributes to the expansion of their academic horizons. Direct experience works as a stimulus for their creativity, enhancing skills such as critical thinking and problem-solving \citep{stein2014stimulating}.

\subsection{RQ3. Do teachers currently encourage the use of AI as a resource to support lectures or assignments?}
Despite the educational potential of AI, 90\% of students indicated that their teachers do not encourage them to use AI technologies as support tools in their classes. It points to a significant gap between students' interest and willingness to use AI and support or promotion by educators. The lack of incentive on the part of teachers suggests an opportunity for improvement in the integration of AI in teaching, highlighting the need to educate and train teachers in these emerging technologies to maximize their potential in the classroom.

\subsection{RQ4. Can the use of generative AI help with learning in other disciplines?}
The use of generative AI proves useful for learning in other disciplines. Students use AI to explore deeper into various topics, design engaging learning activities, and enhance their understanding of content. This suggests that it can be a valuable tool for enhancing and supplementing learning in a wide range of subjects and topics. In this study, it was observed that students, to create more fine-tuned prompts, researched other sources, learning from them in an interactive, no-cost, game-like manner. This practice not only enriches their knowledge in the specific discipline but also fosters research skills and synthesis of information playfully and effectively. In addition, the use of generative AI allows personalizing the study and adapting resources and activities to the individual needs and interests of each student, which contributes to deeper and more personalized learning.

\section{Conclusions}
\label{concl}

The AI rapid learning methodology proposed in this study has proven to be effective in motivating and fostering participation and collaboration among students, even in uncontrolled environments where they do not know each other. The proposed activities also stimulate interest in STEM-related careers, enhancing creativity and highlighting the role of AI as a support tool, while remarking that not all tasks can be performed solely by AI. As it has been observed the sample in gender is unbalanced and given that this type of methodologies allow to stimulate interest, it would be interesting to be able to offer this type of activities in educational centers in order that there is a gender equality in the interest of pursuing technical careers.

Most of the participants indicate that they would consider using generative AI as a support tool in their classes, covering various areas of knowledge to strengthen the understanding of the content. However, 90\% of them indicate that their teachers do not encourage them to use these technologies. This situation represents an opportunity for significant improvement. It would be beneficial for teachers of different subjects to be trained in these technologies to integrate them effectively into their teaching, thus allowing them to take full advantage of the educational potential of AI in the classroom. Through this formative framework, students have a direct understanding of the positive potential of AI for the future but also understand the associated risks themselves, indicating a self-directed learning curve regarding the ethical considerations of AI use.

This methodology appears to have a significant impact on students' perceptions and interest in AI. Encouraging active participation in classes by maintaining student interest, promoting technological collaboration, and enhancing students' experience with AI tools are key areas for improving education. This study highlights the importance of integrating such methodologies to not only educate but also prepare students for responsible and ethical use of AI in their future activities.


\bibliographystyle{cas-model2-names}

\bibliography{cas-refs}

\begin{thebibliography}{66}
\expandafter\ifx\csname natexlab\endcsname\relax\def\natexlab#1{#1}\fi
\providecommand{\url}[1]{\texttt{#1}}
\providecommand{\href}[2]{#2}
\providecommand{\path}[1]{#1}
\providecommand{\DOIprefix}{doi:}
\providecommand{\ArXivprefix}{arXiv:}
\providecommand{\URLprefix}{URL: }
\providecommand{\Pubmedprefix}{pmid:}
\providecommand{\doi}[1]{\href{http://dx.doi.org/#1}{\path{#1}}}
\providecommand{\Pubmed}[1]{\href{pmid:#1}{\path{#1}}}
\providecommand{\bibinfo}[2]{#2}
\ifx\xfnm\relax \def\xfnm[#1]{\unskip,\space#1}\fi
\bibitem[{Achiam et~al.(2023)Achiam, Adler, Agarwal, Ahmad, Akkaya, Aleman, Almeida, Altenschmidt, Altman, Anadkat et~al.}]{achiam2023gpt}
\bibinfo{author}{Achiam, J.}, \bibinfo{author}{Adler, S.}, \bibinfo{author}{Agarwal, S.}, \bibinfo{author}{Ahmad, L.}, \bibinfo{author}{Akkaya, I.}, \bibinfo{author}{Aleman, F.L.}, \bibinfo{author}{Almeida, D.}, \bibinfo{author}{Altenschmidt, J.}, \bibinfo{author}{Altman, S.}, \bibinfo{author}{Anadkat, S.}, et~al., \bibinfo{year}{2023}.
\newblock \bibinfo{title}{Gpt-4 technical report}.
\newblock \bibinfo{journal}{arXiv preprint arXiv:2303.08774} .
\bibitem[{Alasadi and Baiz(2023)}]{alasadi2023generative}
\bibinfo{author}{Alasadi, E.A.}, \bibinfo{author}{Baiz, C.R.}, \bibinfo{year}{2023}.
\newblock \bibinfo{title}{Generative ai in education and research: Opportunities, concerns, and solutions}.
\newblock \bibinfo{journal}{Journal of Chemical Education} \bibinfo{volume}{100}, \bibinfo{pages}{2965--2971}.
\bibitem[{Ali et~al.(2024)Ali, Ravi, Williams, DiPaola and Breazeal}]{ali2024constructing}
\bibinfo{author}{Ali, S.}, \bibinfo{author}{Ravi, P.}, \bibinfo{author}{Williams, R.}, \bibinfo{author}{DiPaola, D.}, \bibinfo{author}{Breazeal, C.}, \bibinfo{year}{2024}.
\newblock \bibinfo{title}{Constructing dreams using generative ai}, in: \bibinfo{booktitle}{Proceedings of the AAAI Conference on Artificial Intelligence}, pp. \bibinfo{pages}{23268--23275}.
\bibitem[{Baidoo-Anu and Ansah(2023)}]{baidoo2023education}
\bibinfo{author}{Baidoo-Anu, D.}, \bibinfo{author}{Ansah, L.O.}, \bibinfo{year}{2023}.
\newblock \bibinfo{title}{Education in the era of generative artificial intelligence (ai): Understanding the potential benefits of chatgpt in promoting teaching and learning}.
\newblock \bibinfo{journal}{Journal of AI} \bibinfo{volume}{7}, \bibinfo{pages}{52--62}.
\bibitem[{Bartsch and Viehoff(2010)}]{bartsch2010use}
\bibinfo{author}{Bartsch, A.}, \bibinfo{author}{Viehoff, R.}, \bibinfo{year}{2010}.
\newblock \bibinfo{title}{The use of media entertainment and emotional gratification}.
\newblock \bibinfo{journal}{Procedia-Social and Behavioral Sciences} \bibinfo{volume}{5}, \bibinfo{pages}{2247--2255}.
\bibitem[{Borji(2022)}]{borji2022generated}
\bibinfo{author}{Borji, A.}, \bibinfo{year}{2022}.
\newblock \bibinfo{title}{Generated faces in the wild: Quantitative comparison of stable diffusion, midjourney and dall-e 2}.
\newblock \bibinfo{journal}{arXiv preprint arXiv:2210.00586} .
\bibitem[{Briot et~al.(2017)Briot, Hadjeres and Pachet}]{briot2017deep}
\bibinfo{author}{Briot, J.P.}, \bibinfo{author}{Hadjeres, G.}, \bibinfo{author}{Pachet, F.D.}, \bibinfo{year}{2017}.
\newblock \bibinfo{title}{Deep learning techniques for music generation--a survey}.
\newblock \bibinfo{journal}{arXiv preprint arXiv:1709.01620} .
\bibitem[{Cao et~al.(2024)Cao, Tan, Gao, Xu, Chen, Heng and Li}]{cao2024survey}
\bibinfo{author}{Cao, H.}, \bibinfo{author}{Tan, C.}, \bibinfo{author}{Gao, Z.}, \bibinfo{author}{Xu, Y.}, \bibinfo{author}{Chen, G.}, \bibinfo{author}{Heng, P.A.}, \bibinfo{author}{Li, S.Z.}, \bibinfo{year}{2024}.
\newblock \bibinfo{title}{A survey on generative diffusion models}.
\newblock \bibinfo{journal}{IEEE Transactions on Knowledge and Data Engineering} .
\bibitem[{CC.MM()}]{4eso}
CC.MM, \bibinfo{year}{2024}.
\newblock \URLprefix \url{https://www.comunidad.madrid/servicios/educacion/programa-4o-esoempresa}. \bibinfo{note}{programa de 4º ESO + Empresa de la comunidad de Madrid, Consejería de Educación, Comunidad de Madrid.}
\bibitem[{Chiu(2023)}]{chiu2023impact}
\bibinfo{author}{Chiu, T.K.}, \bibinfo{year}{2023}.
\newblock \bibinfo{title}{The impact of generative ai (genai) on practices, policies and research direction in education: A case of chatgpt and midjourney}.
\newblock \bibinfo{journal}{Interactive Learning Environments} , \bibinfo{pages}{1--17}.
\bibitem[{Dhariwal and Nichol(2021)}]{dhariwal2021diffusion}
\bibinfo{author}{Dhariwal, P.}, \bibinfo{author}{Nichol, A.}, \bibinfo{year}{2021}.
\newblock \bibinfo{title}{Diffusion models beat gans on image synthesis}.
\newblock \bibinfo{journal}{Advances in neural information processing systems} \bibinfo{volume}{34}, \bibinfo{pages}{8780--8794}.
\bibitem[{Dong et~al.(2018)Dong, Hsiao, Yang and Yang}]{dong2018musegan}
\bibinfo{author}{Dong, H.W.}, \bibinfo{author}{Hsiao, W.Y.}, \bibinfo{author}{Yang, L.C.}, \bibinfo{author}{Yang, Y.H.}, \bibinfo{year}{2018}.
\newblock \bibinfo{title}{Musegan: Multi-track sequential generative adversarial networks for symbolic music generation and accompaniment}, in: \bibinfo{booktitle}{Proceedings of the AAAI Conference on Artificial Intelligence}.
\bibitem[{Famaye et~al.(2023)Famaye, Adisa and Irgens}]{famaye2023ban}
\bibinfo{author}{Famaye, T.}, \bibinfo{author}{Adisa, I.O.}, \bibinfo{author}{Irgens, G.A.}, \bibinfo{year}{2023}.
\newblock \bibinfo{title}{To ban or embrace: Students’ perceptions towards adopting advanced ai chatbots in schools}, in: \bibinfo{booktitle}{International Conference on Quantitative Ethnography}, \bibinfo{organization}{Springer}. pp. \bibinfo{pages}{140--154}.
\bibitem[{Ferguson and Roofe(2020)}]{ferguson2020sdg}
\bibinfo{author}{Ferguson, T.}, \bibinfo{author}{Roofe, C.G.}, \bibinfo{year}{2020}.
\newblock \bibinfo{title}{Sdg 4 in higher education: Challenges and opportunities}.
\newblock \bibinfo{journal}{International Journal of Sustainability in Higher Education} \bibinfo{volume}{21}, \bibinfo{pages}{959--975}.
\bibitem[{Ferriz-Valero et~al.(2020)Ferriz-Valero, {\O}sterlie, Garc{\'\i}a~Mart{\'\i}nez and Garc{\'\i}a-Ja{\'e}n}]{ferriz2020gamification}
\bibinfo{author}{Ferriz-Valero, A.}, \bibinfo{author}{{\O}sterlie, O.}, \bibinfo{author}{Garc{\'\i}a~Mart{\'\i}nez, S.}, \bibinfo{author}{Garc{\'\i}a-Ja{\'e}n, M.}, \bibinfo{year}{2020}.
\newblock \bibinfo{title}{Gamification in physical education: Evaluation of impact on motivation and academic performance within higher education}.
\newblock \bibinfo{journal}{International Journal of Environmental Research and Public Health} \bibinfo{volume}{17}, \bibinfo{pages}{4465}.
\bibitem[{Gan and Khoo(2024)}]{gan2024barriers}
\bibinfo{author}{Gan, K.L.}, \bibinfo{author}{Khoo, T.B.}, \bibinfo{year}{2024}.
\newblock \bibinfo{title}{Barriers to education for children with neurodisabilities in a developing country}.
\newblock \bibinfo{journal}{Cureus} \bibinfo{volume}{16}.
\bibitem[{Gozalo-Brizuela and Garrido-Merchan(2023)}]{gozalo2023chatgpt}
\bibinfo{author}{Gozalo-Brizuela, R.}, \bibinfo{author}{Garrido-Merchan, E.C.}, \bibinfo{year}{2023}.
\newblock \bibinfo{title}{Chatgpt is not all you need. a state of the art review of large generative ai models}.
\newblock \bibinfo{journal}{arXiv preprint arXiv:2301.04655} .
\bibitem[{Ha and Eck(2017)}]{ha2017neural}
\bibinfo{author}{Ha, D.}, \bibinfo{author}{Eck, D.}, \bibinfo{year}{2017}.
\newblock \bibinfo{title}{A neural representation of sketch drawings}.
\newblock \bibinfo{journal}{arXiv preprint arXiv:1704.03477} \URLprefix \url{https://www.autodraw.com/}.
\bibitem[{Han and Cai(2023)}]{han2023design}
\bibinfo{author}{Han, A.}, \bibinfo{author}{Cai, Z.}, \bibinfo{year}{2023}.
\newblock \bibinfo{title}{Design implications of generative ai systems for visual storytelling for young learners}, in: \bibinfo{booktitle}{Proceedings of the 22nd Annual ACM Interaction Design and Children Conference}, pp. \bibinfo{pages}{470--474}.
\bibitem[{Henry et~al.(2003)Henry, Osborne and Salzberger-Wittenberg}]{henry2003emotional}
\bibinfo{author}{Henry, G.}, \bibinfo{author}{Osborne, E.}, \bibinfo{author}{Salzberger-Wittenberg, I.}, \bibinfo{year}{2003}.
\newblock \bibinfo{title}{The emotional experience of learning and teaching}.
\newblock \bibinfo{publisher}{Routledge}.
\bibitem[{Holmes and Tuomi(2022)}]{holmes2022state}
\bibinfo{author}{Holmes, W.}, \bibinfo{author}{Tuomi, I.}, \bibinfo{year}{2022}.
\newblock \bibinfo{title}{State of the art and practice in ai in education}.
\newblock \bibinfo{journal}{European Journal of Education} \bibinfo{volume}{57}, \bibinfo{pages}{542--570}.
\bibitem[{Huang et~al.(2021)Huang, Saleh and Liu}]{huang2021review}
\bibinfo{author}{Huang, J.}, \bibinfo{author}{Saleh, S.}, \bibinfo{author}{Liu, Y.}, \bibinfo{year}{2021}.
\newblock \bibinfo{title}{A review on artificial intelligence in education}.
\newblock \bibinfo{journal}{Academic Journal of Interdisciplinary Studies} \bibinfo{volume}{10}.
\bibitem[{Johnson(1981)}]{johnson1981student}
\bibinfo{author}{Johnson, D.W.}, \bibinfo{year}{1981}.
\newblock \bibinfo{title}{Student-student interaction: The neglected variable in education}.
\newblock \bibinfo{journal}{Educational researcher} \bibinfo{volume}{10}, \bibinfo{pages}{5--10}.
\bibitem[{Kamposiori et~al.(2022)Kamposiori, Warwick and Mahony}]{kamposiori2022embedding}
\bibinfo{author}{Kamposiori, C.}, \bibinfo{author}{Warwick, C.}, \bibinfo{author}{Mahony, S.}, \bibinfo{year}{2022}.
\newblock \bibinfo{title}{Embedding creativity into digital resources: Improving information discovery for art history}.
\newblock \bibinfo{journal}{Digital Scholarship in the Humanities} \bibinfo{volume}{37}, \bibinfo{pages}{469--482}.
\bibitem[{Kansaksiri et~al.(2023)Kansaksiri, Panomkhet and Tantisuwichwong}]{kansaksiri2023smart}
\bibinfo{author}{Kansaksiri, P.}, \bibinfo{author}{Panomkhet, P.}, \bibinfo{author}{Tantisuwichwong, N.}, \bibinfo{year}{2023}.
\newblock \bibinfo{title}{Smart cuisine: Generative recipe \& chatgpt powered nutrition assistance for sustainable cooking}.
\newblock \bibinfo{journal}{Procedia Computer Science} \bibinfo{volume}{225}, \bibinfo{pages}{2028--2036}.
\bibitem[{Karaarslan and Ayd{\i}n(2024)}]{karaarslan2024generate}
\bibinfo{author}{Karaarslan, E.}, \bibinfo{author}{Ayd{\i}n, {\"O}.}, \bibinfo{year}{2024}.
\newblock \bibinfo{title}{Generate impressive videos with text instructions: A review of openai sora, stable diffusion, lumiere and comparable models}.
\newblock \bibinfo{journal}{Authorea Preprints} .
\bibitem[{Karras et~al.(2023)Karras, Holynski, Wang and Kemelmacher-Shlizerman}]{karras2023dreampose}
\bibinfo{author}{Karras, J.}, \bibinfo{author}{Holynski, A.}, \bibinfo{author}{Wang, T.C.}, \bibinfo{author}{Kemelmacher-Shlizerman, I.}, \bibinfo{year}{2023}.
\newblock \bibinfo{title}{Dreampose: Fashion video synthesis with stable diffusion}, in: \bibinfo{booktitle}{Proceedings of the IEEE/CVF International Conference on Computer Vision}, pp. \bibinfo{pages}{22680--22690}.
\bibitem[{Khachatryan et~al.(2023)Khachatryan, Movsisyan, Tadevosyan, Henschel, Wang, Navasardyan and Shi}]{khachatryan2023text2video}
\bibinfo{author}{Khachatryan, L.}, \bibinfo{author}{Movsisyan, A.}, \bibinfo{author}{Tadevosyan, V.}, \bibinfo{author}{Henschel, R.}, \bibinfo{author}{Wang, Z.}, \bibinfo{author}{Navasardyan, S.}, \bibinfo{author}{Shi, H.}, \bibinfo{year}{2023}.
\newblock \bibinfo{title}{Text2video-zero: Text-to-image diffusion models are zero-shot video generators}, in: \bibinfo{booktitle}{Proceedings of the IEEE/CVF International Conference on Computer Vision}, pp. \bibinfo{pages}{15954--15964}.
\bibitem[{Leiker et~al.(2023)Leiker, Gyllen, Eldesouky and Cukurova}]{leiker2023generative}
\bibinfo{author}{Leiker, D.}, \bibinfo{author}{Gyllen, A.R.}, \bibinfo{author}{Eldesouky, I.}, \bibinfo{author}{Cukurova, M.}, \bibinfo{year}{2023}.
\newblock \bibinfo{title}{Generative ai for learning: Investigating the potential of synthetic learning videos}.
\newblock \bibinfo{journal}{arXiv preprint arXiv:2304.03784} .
\bibitem[{Lim et~al.(2023)Lim, Gunasekara, Pallant, Pallant and Pechenkina}]{lim2023generative}
\bibinfo{author}{Lim, W.M.}, \bibinfo{author}{Gunasekara, A.}, \bibinfo{author}{Pallant, J.L.}, \bibinfo{author}{Pallant, J.I.}, \bibinfo{author}{Pechenkina, E.}, \bibinfo{year}{2023}.
\newblock \bibinfo{title}{Generative ai and the future of education: Ragnar{\"o}k or reformation? a paradoxical perspective from management educators}.
\newblock \bibinfo{journal}{The international journal of management education} \bibinfo{volume}{21}, \bibinfo{pages}{100790}.
\bibitem[{Liua et~al.(2021)Liua, Salehb and Huang}]{liua2021artificial}
\bibinfo{author}{Liua, Y.}, \bibinfo{author}{Salehb, S.}, \bibinfo{author}{Huang, J.}, \bibinfo{year}{2021}.
\newblock \bibinfo{title}{Artificial intelligence in promoting teaching and learning transformation in schools}.
\newblock \bibinfo{journal}{Artificial Intelligence} \bibinfo{volume}{15}.
\bibitem[{Magraner and Valero(2016)}]{magraner2016educacion}
\bibinfo{author}{Magraner, J.S.B.}, \bibinfo{author}{Valero, G.B.}, \bibinfo{year}{2016}.
\newblock \bibinfo{title}{Educaci{\'o}n musical y competencia colaborativa: una experiencia con alumnado universitario}.
\newblock \bibinfo{journal}{Magister} \bibinfo{volume}{28}, \bibinfo{pages}{63--70}.
\bibitem[{M{\o}rch and Andersen(2023)}]{morch2023human}
\bibinfo{author}{M{\o}rch, A.I.}, \bibinfo{author}{Andersen, R.}, \bibinfo{year}{2023}.
\newblock \bibinfo{title}{Human-centred ai in education in the age of generative ai tools}.
\newblock \bibinfo{journal}{Proceedings http://ceur-ws. org ISSN} \bibinfo{volume}{1613}, \bibinfo{pages}{0073}.
\bibitem[{Mulang and Putra(2023)}]{mulang2023exploring}
\bibinfo{author}{Mulang, H.}, \bibinfo{author}{Putra, A.H.P.K.}, \bibinfo{year}{2023}.
\newblock \bibinfo{title}{Exploring the implementation of ethical and spiritual values in high school education: A case study in makassar, indonesia}.
\newblock \bibinfo{journal}{Golden Ratio of Social Science and Education} \bibinfo{volume}{3}, \bibinfo{pages}{01--13}.
\bibitem[{Ngo(2023)}]{ngo2023perception}
\bibinfo{author}{Ngo, T.T.A.}, \bibinfo{year}{2023}.
\newblock \bibinfo{title}{The perception by university students of the use of chatgpt in education}.
\newblock \bibinfo{journal}{International Journal of Emerging Technologies in Learning (Online)} \bibinfo{volume}{18}, \bibinfo{pages}{4}.
\bibitem[{Patreon(2022)}]{Deforum_v071}
\bibinfo{author}{Patreon}, \bibinfo{year}{2022}.
\newblock \bibinfo{title}{Deforum stable diffusion (v0.7.1)}.
\newblock \URLprefix \url{https://colab.research.google.com/github/deforum-art/deforum-stable-diffusion/blob/main/Deforum_Stable_Diffusion.ipynb?authuser=1#scrollTo=ByGXyiHZWM_q}.
\bibitem[{Perez et~al.(2020)Perez, Dedden and Goodloe}]{perez2020copilot}
\bibinfo{author}{Perez, I.}, \bibinfo{author}{Dedden, F.}, \bibinfo{author}{Goodloe, A.}, \bibinfo{year}{2020}.
\newblock \bibinfo{title}{Copilot 3}.
\newblock \bibinfo{type}{Technical Report}.
\bibitem[{Pesovski et~al.(2024)Pesovski, Santos, Henriques and Trajkovik}]{pesovski2024generative}
\bibinfo{author}{Pesovski, I.}, \bibinfo{author}{Santos, R.}, \bibinfo{author}{Henriques, R.}, \bibinfo{author}{Trajkovik, V.}, \bibinfo{year}{2024}.
\newblock \bibinfo{title}{Generative ai for customizable learning experiences}.
\newblock \bibinfo{journal}{Sustainability} \bibinfo{volume}{16}, \bibinfo{pages}{3034}.
\bibitem[{Quay(2003)}]{quay2003experience}
\bibinfo{author}{Quay, J.}, \bibinfo{year}{2003}.
\newblock \bibinfo{title}{Experience and participation: Relating theories of learning}.
\newblock \bibinfo{journal}{Journal of Experiential Education} \bibinfo{volume}{26}, \bibinfo{pages}{105--112}.
\bibitem[{Ruiz-Rojas et~al.(2023)Ruiz-Rojas, Acosta-Vargas, De-Moreta-Llovet and Gonzalez-Rodriguez}]{ruiz2023empowering}
\bibinfo{author}{Ruiz-Rojas, L.I.}, \bibinfo{author}{Acosta-Vargas, P.}, \bibinfo{author}{De-Moreta-Llovet, J.}, \bibinfo{author}{Gonzalez-Rodriguez, M.}, \bibinfo{year}{2023}.
\newblock \bibinfo{title}{Empowering education with generative artificial intelligence tools: Approach with an instructional design matrix}.
\newblock \bibinfo{journal}{Sustainability} \bibinfo{volume}{15}, \bibinfo{pages}{11524}.
\bibitem[{Salimans et~al.(2016)Salimans, Goodfellow, Zaremba, Cheung, Radford and Chen}]{salimans2016improved}
\bibinfo{author}{Salimans, T.}, \bibinfo{author}{Goodfellow, I.}, \bibinfo{author}{Zaremba, W.}, \bibinfo{author}{Cheung, V.}, \bibinfo{author}{Radford, A.}, \bibinfo{author}{Chen, X.}, \bibinfo{year}{2016}.
\newblock \bibinfo{title}{Improved techniques for training gans}.
\newblock \bibinfo{journal}{Advances in neural information processing systems} \bibinfo{volume}{29}.
\bibitem[{Schrum et~al.(2020)Schrum, Johnson, Ghuy and Gombolay}]{schrum2020four}
\bibinfo{author}{Schrum, M.L.}, \bibinfo{author}{Johnson, M.}, \bibinfo{author}{Ghuy, M.}, \bibinfo{author}{Gombolay, M.C.}, \bibinfo{year}{2020}.
\newblock \bibinfo{title}{Four years in review: Statistical practices of likert scales in human-robot interaction studies}, in: \bibinfo{booktitle}{Companion of the 2020 ACM/IEEE International Conference on Human-Robot Interaction}, pp. \bibinfo{pages}{43--52}.
\bibitem[{Sdg(2019)}]{sdg2019sustainable}
\bibinfo{author}{Sdg, U.}, \bibinfo{year}{2019}.
\newblock \bibinfo{title}{Sustainable development goals}.
\newblock \bibinfo{journal}{The energy progress report. Tracking SDG} \bibinfo{volume}{7}, \bibinfo{pages}{805--814}.
\bibitem[{Seechaliao(2017)}]{seechaliao2017instructional}
\bibinfo{author}{Seechaliao, T.}, \bibinfo{year}{2017}.
\newblock \bibinfo{title}{Instructional strategies to support creativity and innovation in education.}
\newblock \bibinfo{journal}{Journal of education and learning} \bibinfo{volume}{6}, \bibinfo{pages}{201--208}.
\bibitem[{Sharples(2023)}]{sharples2023towards}
\bibinfo{author}{Sharples, M.}, \bibinfo{year}{2023}.
\newblock \bibinfo{title}{Towards social generative ai for education: theory, practices and ethics}.
\newblock \bibinfo{journal}{Learning: Research and Practice} \bibinfo{volume}{9}, \bibinfo{pages}{159--167}.
\bibitem[{Singh(2023)}]{singh2023review}
\bibinfo{author}{Singh, A.}, \bibinfo{year}{2023}.
\newblock \bibinfo{title}{A review on objective-driven artificial intelligence}.
\newblock \bibinfo{journal}{arXiv preprint arXiv:2308.10135} .
\bibitem[{Stein(2014)}]{stein2014stimulating}
\bibinfo{author}{Stein, M.I.}, \bibinfo{year}{2014}.
\newblock \bibinfo{title}{Stimulating creativity: Individual procedures}.
\newblock \bibinfo{publisher}{Academic Press}.
\bibitem[{Su and Yang(2023)}]{su2023unlocking}
\bibinfo{author}{Su, J.}, \bibinfo{author}{Yang, W.}, \bibinfo{year}{2023}.
\newblock \bibinfo{title}{Unlocking the power of chatgpt: A framework for applying generative ai in education}.
\newblock \bibinfo{journal}{ECNU Review of Education} \bibinfo{volume}{6}, \bibinfo{pages}{355--366}.
\bibitem[{Sulistiobudi and Kadiyono(2023)}]{sulistiobudi2023employability}
\bibinfo{author}{Sulistiobudi, R.A.}, \bibinfo{author}{Kadiyono, A.L.}, \bibinfo{year}{2023}.
\newblock \bibinfo{title}{Employability of students in vocational secondary school: Role of psychological capital and student-parent career congruences}.
\newblock \bibinfo{journal}{Heliyon} \bibinfo{volume}{9}.
\bibitem[{Tan(2008)}]{tan2008entertainment}
\bibinfo{author}{Tan, E.S.H.}, \bibinfo{year}{2008}.
\newblock \bibinfo{title}{Entertainment is emotion: The functional architecture of the entertainment experience}.
\newblock \bibinfo{journal}{Media psychology} \bibinfo{volume}{11}, \bibinfo{pages}{28--51}.
\bibitem[{Team et~al.(2023)Team, Anil, Borgeaud, Wu, Alayrac, Yu, Soricut, Schalkwyk, Dai, Hauth et~al.}]{team2023gemini}
\bibinfo{author}{Team, G.}, \bibinfo{author}{Anil, R.}, \bibinfo{author}{Borgeaud, S.}, \bibinfo{author}{Wu, Y.}, \bibinfo{author}{Alayrac, J.B.}, \bibinfo{author}{Yu, J.}, \bibinfo{author}{Soricut, R.}, \bibinfo{author}{Schalkwyk, J.}, \bibinfo{author}{Dai, A.M.}, \bibinfo{author}{Hauth, A.}, et~al., \bibinfo{year}{2023}.
\newblock \bibinfo{title}{Gemini: a family of highly capable multimodal models}.
\newblock \bibinfo{journal}{arXiv preprint arXiv:2312.11805} .
\bibitem[{Tossell et~al.(2024)Tossell, Tenhundfeld, Momen, Cooley and de~Visser}]{tossell2024student}
\bibinfo{author}{Tossell, C.C.}, \bibinfo{author}{Tenhundfeld, N.L.}, \bibinfo{author}{Momen, A.}, \bibinfo{author}{Cooley, K.}, \bibinfo{author}{de~Visser, E.J.}, \bibinfo{year}{2024}.
\newblock \bibinfo{title}{Student perceptions of chatgpt use in a college essay assignment: Implications for learning, grading, and trust in artificial intelligence}.
\newblock \bibinfo{journal}{IEEE Transactions on Learning Technologies} .
\bibitem[{Touvron et~al.(2023)Touvron, Martin, Stone, Albert, Almahairi, Babaei, Bashlykov, Batra, Bhargava, Bhosale et~al.}]{touvron2023llama}
\bibinfo{author}{Touvron, H.}, \bibinfo{author}{Martin, L.}, \bibinfo{author}{Stone, K.}, \bibinfo{author}{Albert, P.}, \bibinfo{author}{Almahairi, A.}, \bibinfo{author}{Babaei, Y.}, \bibinfo{author}{Bashlykov, N.}, \bibinfo{author}{Batra, S.}, \bibinfo{author}{Bhargava, P.}, \bibinfo{author}{Bhosale, S.}, et~al., \bibinfo{year}{2023}.
\newblock \bibinfo{title}{Llama 2: Open foundation and fine-tuned chat models}.
\newblock \bibinfo{journal}{arXiv preprint arXiv:2307.09288} .
\bibitem[{Vallis et~al.(2023)Vallis, Wilson, Gozman and Buchanan}]{vallis2023student}
\bibinfo{author}{Vallis, C.}, \bibinfo{author}{Wilson, S.}, \bibinfo{author}{Gozman, D.}, \bibinfo{author}{Buchanan, J.}, \bibinfo{year}{2023}.
\newblock \bibinfo{title}{Student perceptions of ai-generated avatars in teaching business ethics: We might not be impressed}.
\newblock \bibinfo{journal}{Postdigital Science and Education} , \bibinfo{pages}{1--19}.
\bibitem[{Vartiainen and Tedre(2023)}]{vartiainen2023using}
\bibinfo{author}{Vartiainen, H.}, \bibinfo{author}{Tedre, M.}, \bibinfo{year}{2023}.
\newblock \bibinfo{title}{Using artificial intelligence in craft education: crafting with text-to-image generative models}.
\newblock \bibinfo{journal}{Digital Creativity} \bibinfo{volume}{34}, \bibinfo{pages}{1--21}.
\bibitem[{Vaswani et~al.(2017)Vaswani, Shazeer, Parmar, Uszkoreit, Jones, Gomez, Kaiser and Polosukhin}]{vaswani2017attention}
\bibinfo{author}{Vaswani, A.}, \bibinfo{author}{Shazeer, N.}, \bibinfo{author}{Parmar, N.}, \bibinfo{author}{Uszkoreit, J.}, \bibinfo{author}{Jones, L.}, \bibinfo{author}{Gomez, A.N.}, \bibinfo{author}{Kaiser, {\L}.}, \bibinfo{author}{Polosukhin, I.}, \bibinfo{year}{2017}.
\newblock \bibinfo{title}{Attention is all you need}.
\newblock \bibinfo{journal}{Advances in neural information processing systems} \bibinfo{volume}{30}.
\bibitem[{Wang et~al.(2024)Wang, Zhao, Liu, Pang, Qin and Wu}]{wang2024review}
\bibinfo{author}{Wang, L.}, \bibinfo{author}{Zhao, Z.}, \bibinfo{author}{Liu, H.}, \bibinfo{author}{Pang, J.}, \bibinfo{author}{Qin, Y.}, \bibinfo{author}{Wu, Q.}, \bibinfo{year}{2024}.
\newblock \bibinfo{title}{A review of intelligent music generation systems}.
\newblock \bibinfo{journal}{Neural Computing and Applications} , \bibinfo{pages}{1--21}.
\bibitem[{White et~al.(2023)White, Fu, Hays, Sandborn, Olea, Gilbert, Elnashar, Spencer-Smith and Schmidt}]{white2023prompt}
\bibinfo{author}{White, J.}, \bibinfo{author}{Fu, Q.}, \bibinfo{author}{Hays, S.}, \bibinfo{author}{Sandborn, M.}, \bibinfo{author}{Olea, C.}, \bibinfo{author}{Gilbert, H.}, \bibinfo{author}{Elnashar, A.}, \bibinfo{author}{Spencer-Smith, J.}, \bibinfo{author}{Schmidt, D.C.}, \bibinfo{year}{2023}.
\newblock \bibinfo{title}{A prompt pattern catalog to enhance prompt engineering with chatgpt}.
\newblock \bibinfo{journal}{arXiv preprint arXiv:2302.11382} .
\bibitem[{Wiedermann and von Eye(2013)}]{wiedermann2013robustness}
\bibinfo{author}{Wiedermann, W.}, \bibinfo{author}{von Eye, A.}, \bibinfo{year}{2013}.
\newblock \bibinfo{title}{Robustness and power of the parametric t test and the nonparametric wilcoxon test under non-independence of observations}.
\newblock \bibinfo{journal}{Psychological Test and Assessment Modeling} \bibinfo{volume}{55}, \bibinfo{pages}{39--61}.
\bibitem[{Wolf et~al.(2019)Wolf, Debut, Sanh, Chaumond, Delangue, Moi, Cistac, Rault, Louf, Funtowicz et~al.}]{wolf2019huggingface}
\bibinfo{author}{Wolf, T.}, \bibinfo{author}{Debut, L.}, \bibinfo{author}{Sanh, V.}, \bibinfo{author}{Chaumond, J.}, \bibinfo{author}{Delangue, C.}, \bibinfo{author}{Moi, A.}, \bibinfo{author}{Cistac, P.}, \bibinfo{author}{Rault, T.}, \bibinfo{author}{Louf, R.}, \bibinfo{author}{Funtowicz, M.}, et~al., \bibinfo{year}{2019}.
\newblock \bibinfo{title}{Huggingface's transformers: State-of-the-art natural language processing}.
\newblock \bibinfo{journal}{arXiv preprint arXiv:1910.03771} \URLprefix \url{https://xenova-doodle-dash.static.hf.space/index.html?ref=futuretools.io}.
\bibitem[{Wu et~al.(2023)Wu, Ge, Wang, Lei, Gu, Shi, Hsu, Shan, Qie and Shou}]{wu2023tune}
\bibinfo{author}{Wu, J.Z.}, \bibinfo{author}{Ge, Y.}, \bibinfo{author}{Wang, X.}, \bibinfo{author}{Lei, S.W.}, \bibinfo{author}{Gu, Y.}, \bibinfo{author}{Shi, Y.}, \bibinfo{author}{Hsu, W.}, \bibinfo{author}{Shan, Y.}, \bibinfo{author}{Qie, X.}, \bibinfo{author}{Shou, M.Z.}, \bibinfo{year}{2023}.
\newblock \bibinfo{title}{Tune-a-video: One-shot tuning of image diffusion models for text-to-video generation}, in: \bibinfo{booktitle}{Proceedings of the IEEE/CVF International Conference on Computer Vision}, pp. \bibinfo{pages}{7623--7633}.
\bibitem[{Yang et~al.(2023)Yang, Ogata and Matsui}]{yang2023guest}
\bibinfo{author}{Yang, S.J.}, \bibinfo{author}{Ogata, H.}, \bibinfo{author}{Matsui, T.}, \bibinfo{year}{2023}.
\newblock \bibinfo{title}{Guest editorial: Human-centered ai in education: Augment human intelligence with machine intelligence} .
\bibitem[{Yang et~al.(2021)Yang, Ogata, Matsui and Chen}]{yang2021human}
\bibinfo{author}{Yang, S.J.}, \bibinfo{author}{Ogata, H.}, \bibinfo{author}{Matsui, T.}, \bibinfo{author}{Chen, N.S.}, \bibinfo{year}{2021}.
\newblock \bibinfo{title}{Human-centered artificial intelligence in education: Seeing the invisible through the visible}.
\newblock \bibinfo{journal}{Computers and Education: Artificial Intelligence} \bibinfo{volume}{2}, \bibinfo{pages}{100008}.
\bibitem[{Zaman(2023)}]{zaman2023transforming}
\bibinfo{author}{Zaman, B.U.}, \bibinfo{year}{2023}.
\newblock \bibinfo{title}{Transforming education through ai, benefits, risks, and ethical considerations}.
\newblock \bibinfo{journal}{Authorea Preprints} .
\bibitem[{Zastudil et~al.(2023)Zastudil, Rogalska, Kapp, Vaughn and MacNeil}]{zastudil2023generative}
\bibinfo{author}{Zastudil, C.}, \bibinfo{author}{Rogalska, M.}, \bibinfo{author}{Kapp, C.}, \bibinfo{author}{Vaughn, J.}, \bibinfo{author}{MacNeil, S.}, \bibinfo{year}{2023}.
\newblock \bibinfo{title}{Generative ai in computing education: Perspectives of students and instructors}, in: \bibinfo{booktitle}{2023 IEEE Frontiers in Education Conference (FIE)}, \bibinfo{organization}{IEEE}. pp. \bibinfo{pages}{1--9}.
\bibitem[{Țal{\u{a}} et~al.(2024)Țal{\u{a}}, M{\"u}ller, N{\u{a}}stase, Gheorghe et~al.}]{țalua2024exploring}
\bibinfo{author}{Țal{\u{a}}, M.L.}, \bibinfo{author}{M{\"u}ller, C.N.}, \bibinfo{author}{N{\u{a}}stase, I.A.}, \bibinfo{author}{Gheorghe, G.}, et~al., \bibinfo{year}{2024}.
\newblock \bibinfo{title}{Exploring university students' perceptions of generative artificial intelligence in education}.
\newblock \bibinfo{journal}{Amfiteatru Economic Journal} \bibinfo{volume}{26}, \bibinfo{pages}{71--88}.

\end{thebibliography}

\end{document}